# A Survey of Deep Learning Techniques for the Analysis of COVID-19 and their usability for Detecting Omicron

**Asifullah Khan**[1,2,3*]**, Saddam Hussain Khan**[1,2,4]**, Mahrukh Saif**[1]**, Asiya Batool**[1]**,
Anabia Sohail**[1,2,5] **and Muhammad Waleed Khan**[1,2]

[1]Pattern Recognition Lab, Department of Computer & Information Sciences, Pakistan Institute of Engineering & Applied Sciences, Nilore, Islamabad 45650, Pakistan

[2]PIEAS Artificial Intelligence Center (PAIC), Pakistan Institute of Engineering & Applied Sciences, Nilore, Islamabad 45650, Pakistan

[3]Center for Mathematical Sciences, Pakistan Institute of Engineering & Applied Sciences, Nilore, Islamabad 45650, Pakistan

[4]Department of Computer Systems Engineering, University of Engineering and Applied Sciences, Swat 19060, Pakistan

[5]Computer Science Department, Air University, E-9, Islamabad, Pakistan

**Abstract**

The Coronavirus (COVID-19) outbreak in December 2019 has become an ongoing threat to humans worldwide, creating a health crisis that infected millions of lives, as well as devastating the global economy. Deep learning (DL) techniques have proved helpful in analysis and delineation of infectious regions in radiological images in a timely manner. This paper makes an in-depth survey of DL techniques and draws a taxonomy based on diagnostic strategies and learning approaches. DL techniques are systematically categorized into classification, segmentation, and multi-stage approaches for COVID-19 diagnosis at image and region level analysis. Each category includes pre-trained and custom-made Convolutional Neural Network architectures for detecting COVID-19 infection in radiographic imaging modalities; X-Ray, and Computer Tomography (CT). Furthermore, a discussion is made on challenges in developing diagnostic techniques such as cross-platform interoperability and examining imaging modality. Similarly, a review of the various methodologies and performance measures used in these techniques is also presented. This survey provides an insight into the promising areas of research in DL for analyzing radiographic images, and further accelerates the research in designing customized DL based diagnostic tools for effectively dealing with new variants of COVID-19 and emerging challenges.

Corresponding Author: (asif@pieas.edu.pk)

**List of Abbreviations**

| | |
|---|---|
| COVID-19 | Coronavirus Disease |
| AI | Artificial Intelligence |
| RT-PCR | Reverse Transcript-Polymerase Chain Reaction |
| DL | Deep Learning |
| CT | Computed Tomography |
| ML | Machine Learning |
| TL | Transfer Learning |
| CNN | Convolutional Neural Network |
| CV | Cross-Validation |
| ACC | Accuracy |
| PRE | Precision |
| SEN | Recall or Sensitivity |
| SPEC | Specificity |
| F1 | F1-Score |
| AUC | Area Under the Curve |
| MCC | Matthews Correlation Coefficient |
| PPV | Positive Prediction Value |
| NPV | Negative Prediction Value |
| DS | Dice Similarity |
| IOU | Intersection over Union |
| CAM | Class Activation Map |
| HD | Hausdorff Distance |
| DRO | Distributionally Robust Optimization |
| Sα | Structure Measure |
| EAM | Enhance Alignment Measure |
| FNR | False Negative Rate |
| MAE | Mean Absolute Error |
| FPR | False Positive Rate |
| FMI | Fowlkes–Mallows Index |
| iHU | Infection Hounsfield Unit |
| POI | Portion of Infection |
| GM | Global Mean |
| NSD | Normalized Surface Dice |
| SIRM | Italian Society of Medical and Interventional Radiology |

## 1  Introduction

Coronavirus Disease (COVID-19) belongs to a family of viruses that are around for decades, resulting in an epidemic or pandemic [1], [2]. The world has witnessed thousands of deaths from the current pandemic, namely SARS-COV-2 and many other previous pandemics like MERS-COV, SARS-COV, and Ebolavirus Rotavirus, Marburg virus, Dengue, etc. The current SARS-COV-2, however, is a novel zoonotic ailment. Due to this pandemic's novelty, humans have no innate immunity against its spread [3], [4]. At the time of writing of this article, i.e., April 2022, the current COVID-19 pandemic has infected over 487 million people worldwide, taking more than 6 million lives across the globe. The number of recovered cases is around 422 million [5]. The most common symptoms include highly persistent fever, fatigue, myalgia, dyspnea, headache, dry cough, acute respiratory illness, organ failure, all of which can become fatal in the most severe cases [6]–[11] . In some cases, a person infected with COVID-19 may be asymptomatic, posing an even greater threat to human lives [12].

COVID-19 in most severe cases affects the lungs which can result in Pneumonia. For diagnosing whether the lung is affected by viral, bacterial or COVID-19 Pneumonia, radiographic appearance of lungs such as a chest X-Ray (CX-Ray), CT scan, ultrasound and other can be used [13]–[15]. In a COVID-19 affected lung, the patterns of haziness such as ground-glass opacities (GGO) and linear opacities can be seen [16]. Diagnosing COVID-19 can also be performed using classical laboratory approaches that involve a series of reverse transcript -polymerase chain reaction tests (RT-PCR). The RT-PCR consumes more time and sometimes generates false negative [17]. However, COVID-19 infection can be quickly identified using radiographic images of the lungs. This can lead to early isolation of the infected person, and thus preventing its spread. However, radiological images will not prove helpful if the virus has not yet infected the lungs. In that case, RT-PCR proves effective since it takes sample from nose or throat, where the virus initially gathers and spreads further [18], [19]. Examples of normal and COVID-19 affected samples from CX-Ray and CT are shown in Fig. 1.

Two major areas of Artificial Intelligence (AI), i.e., Machine Learning (ML) and Deep Learning (DL) have been rigorously used in detecting COVID-19 positive cases. DL methods generally aim at learning hierarchical features from the data. This concept of hierarchal feature learning allows the DL methods to efficiently tackle complicated patterns. COVID-19 infection displays specific patterns like GGO, consolidation, pleural effusion, bilateral lung involvement etc. in radiological images [20]. Using different DL architectures, these COVID-19 specific patterns can be identified [21]. In the context of COVID-19 detection, the DL models reported higher sensitivity and specificity values, and hence are considered more accurate. Additionally, it can also decrease negative error and false positive rate, providing medical specialists and radiologists with a quick, economical and accurate diagnostic

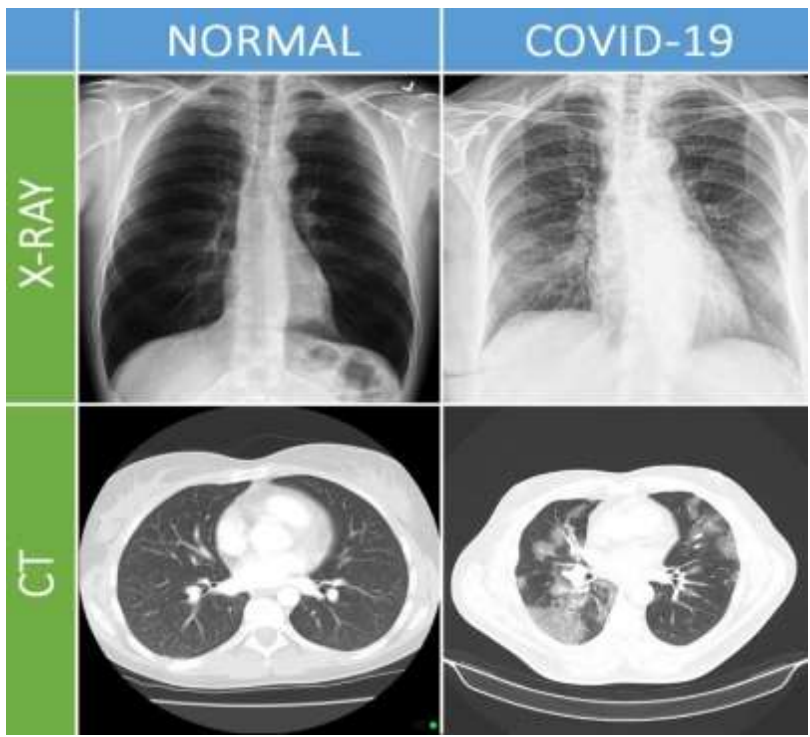

**Figure 1.** Example of normal and COVID-19 affected samples from CX-Ray and CT modalities [22]

System [23], [24]. The DL model can be built and trained entirely from scratch to learn features or it can use the pre-trained model to save time and fine-tune it to perform COVID-19 analysis related tasks.

The main intention behind this survey is to analyze the challenges confronted while controlling COVID-19 spread and the effectiveness of the usage of DL techniques and radiological images. This survey paper will help the researcher for quick insights regarding the efficiency of several CNN models utilized during COVID-19 infection Classification, Segmentation and multi-stage analysis. These analyses give detailed insight into COVID-19 infected radiological images and can help the radiologist to make better decisions for combating the disease. This paper also presents a taxonomy of DL methods used to detect COVID-19 in radiographic images and summarizes the recent relevant literature under this taxonomy. The review is organized into classification, segmentation and multi-stage approach that includes Transfer Learning (TL) based fine-tuned and custom-made CNNs for analyzing infection in radiological images, X-Ray and CT. In the data collection stage of our study, we collected several relevant research articles. Among these, 20 used classification techniques, 20 used segmentation techniques, and 10 research papers used multi-stage approaches.

The rest of the paper is structured as follows: Section 2 discusses the characteristics of radiographic data. Section 3 describes the strategies of COVID-19 detection which include image-level detection and region-level detection. Section 4 explains the hierarchy terms; including classification, segmentation, multi-stage approach, TL and custom architecture. The data sources and general performance metrics used by researchers are highlighted in Section 5. Section 6 analyzes the performance of well-established classification, segmentation and multi-stage techniques, and discusses their contribution and limitations. Section 7 mentions the recent COVID-19 commercial and non-commercial web predictors/software that utilize the DL models. Section 8 presents the challenges related to the dataset collection, implementation of DL methods, and the clinical approval of these techniques. A direction for conducting quality oriented future research and better clinical relevance is presented in Section 9. Finally, the paper is concluded in Section 10. An overview of the flow of the paper is shown is Fig. 2.

## 2  Overview of Radiographic Data and its Characteristics

### 2.1  X-Ray Characteristics

X-Ray, an electromagnetic wave, is commonly used for medical imaging of various organs such as bones and lungs [25]. CX-Ray based techniques provide noninvasive diagnostics of various diseases. A two-dimensional contrast image, also known as a radiograph, is produced due to the tissue's difference in the attenuation of CX-Rays. The CX-Rays are normally captured at various positions and angles for a patient relative to the source and detector panel. Radiographic analysis of CX-Rays is considered one of the quickest and fundamental detection techniques, available at a low cost worldwide.

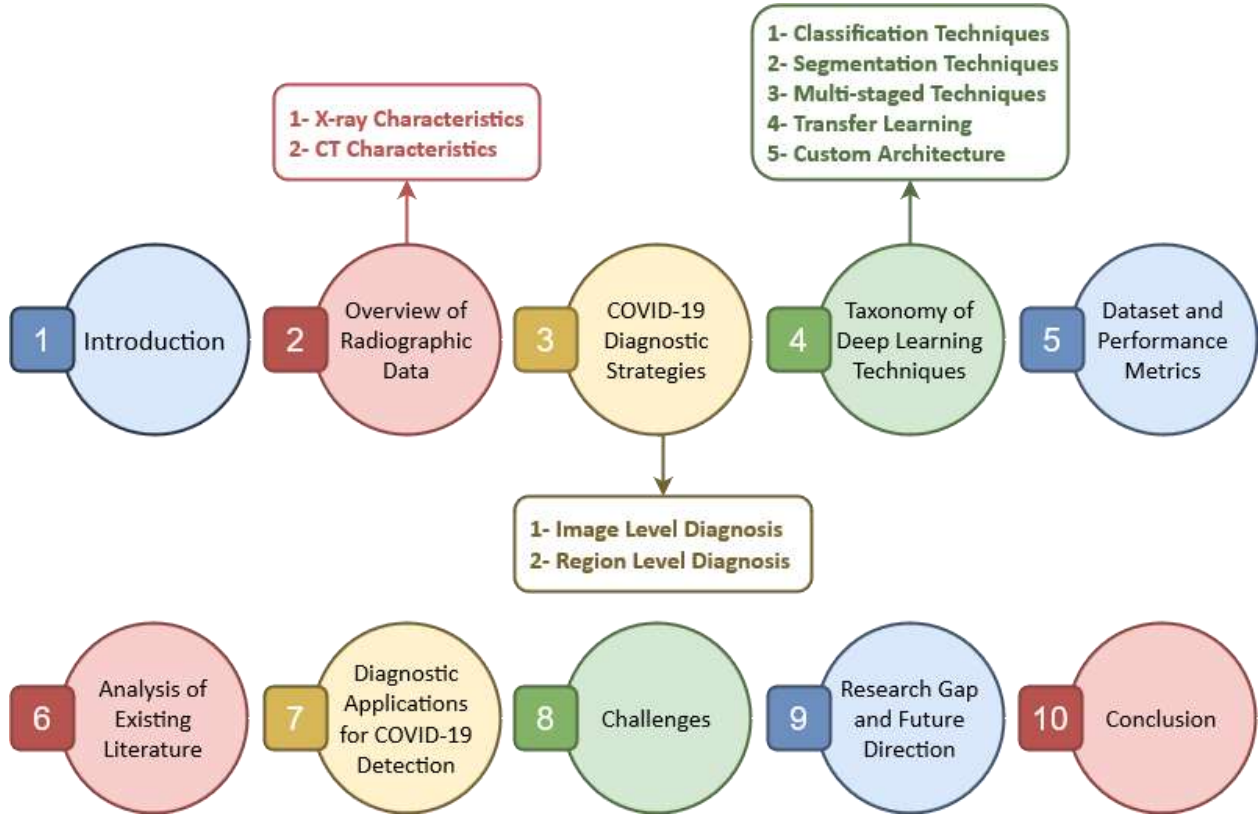

**Figure 2.** Overview of the Proposed Survey.

## 2.2 CT Characteristics

CT, a more sophisticated imaging technique, can also be used to detect small-scale variations in the structure of internal organs [26]. This technique uses three-dimensional computer vision technology in addition to an X-Ray for the detailed analysis of organs. Although both CT scans and X-Ray capture images of internal body structures, the images tend to overlap in the case of traditional X-Rays. On the other hand, in the case of a CT scan, the overlapping issues are mostly resolved, resulting in a clear understanding of the organ under observation.

## 3  COVID-19 Diagnostic Strategies

An overview of the workflow and the methods reviewed for COVID-19 medical image analysis are shown in Fig. 3. Techniques based on both image and region level diagnoses are discussed.

### 3.1 Image Level Diagnosis

This category of method performs image level prediction that involves assigning a label to the entire radiological image of COVID-19 infected patient. These techniques can be considered for initial COVID-19 suspect screening, appropriate diagnosis, and isolation of COVID-infected patients from healthy individuals to control the transmission of the pandemic. DL classification models have utilized medical image-based analysis strategies for disease diagnosis and prognosis [25]. Image level diagnosis assigns binary class labels (separating COVID-19 infected from Normal images) or multi-class labels (separating COVID-19 infected from Normal, viral Pneumonia, bacterial Pneumonia, etc.).

### 3.2 Region Level Diagnosis

Contrary to image level diagnosis, region level diagnosis makes disease prediction by assigning labels to small patches or segmented regions of the radiological image. Every pixel in the image is assigned a category, either "COVID-19" or "other," to demarcate the region of interest [25]. This can provide detailed insight into COVID-19 infection spread and the distinct patterns formed. The region-level diagnosis of COVID-19 images includes DL-based instance and semantic segmentation techniques, and a DL-based multi-stage approach involving both classification and segmentation.

## 4  Taxonomy of DL Techniques for COVID-19 Diagnosis

The manual analysis of COVID-19 infected images or regions requires trained radiologists. However, with the daily increase of COVID-19 cases and other lung diseases, the availability of a sufficient number of radiologists puts a considerable burden on medical care. Therefore, DL based schemes have been employed for automatic COVID-19 diagnosis. The taxonomy of the DL schemes is shown in Fig. 4.

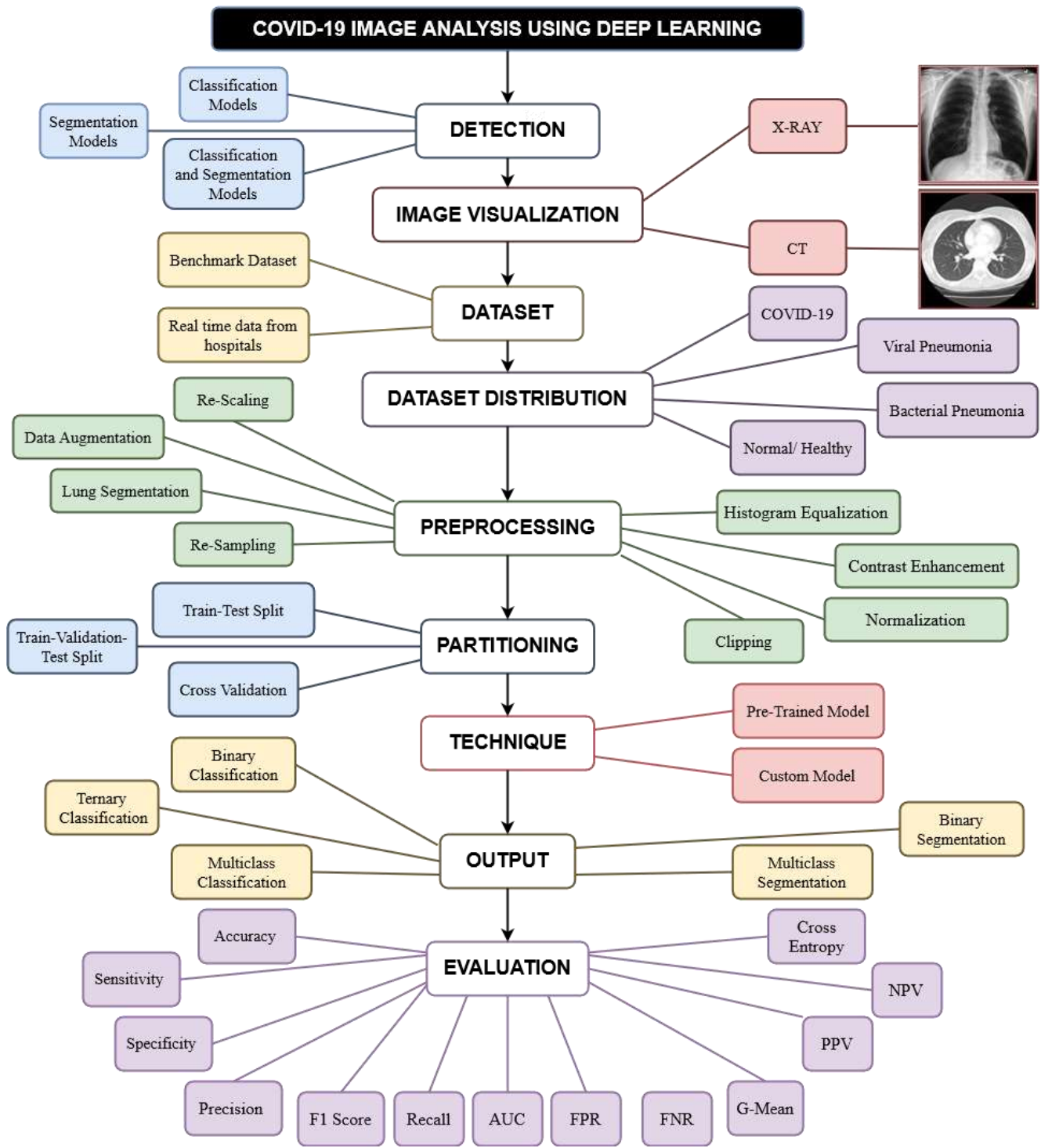

**Figure 3.** Workflow, DL-based techniques. and performance metrics for COVID-19 medical image analysis

## 4.1 Classification Techniques

Classification is a technique of assigning categories to data entities. These techniques have been adapted for the screening of COVID-19 infection in radiological images. These techniques label radiological image data into different categories, including COVID-19 infected, viral and bacterial Pneumonia, Normal, etc. Classification can contribute to efficient detection and timely control and cure of COVID-19. Meticulous analysis of X-Ray and CT images require radiology experts and is time-consuming. DL methods can provide a quick solution by automating the detection process of COVID-19 infected radiological images. This will save time for medical professionals to attend critical COVID-19 patients and save more lives.

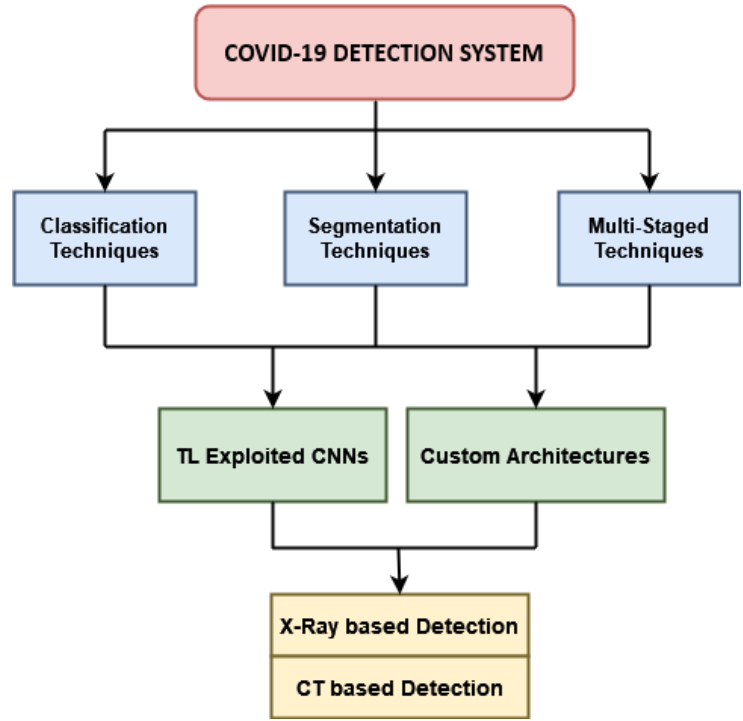

**Figure 4.** Taxonomy of COVID-19 Detection System

## 4.2 Segmentation Techniques

Segmentation of radiological images of COVID-19 infected patients assists in diagnosing and quantifying the severity of the disease (mild, medium or serve stages) [27]. CT scan is a more preferred imaging technique in the segmentation task than X-Ray as it provides an excellent 3-dimensional view of the lung regions [28]. Manual segmentation by an expert radiologist can take hours to accurately segment the infected lesions from CT scans. Hence there is a need for designing automated methods for the segmentation of lesions analyzed in the CT lungs images of COVID-19 patients. Recently, a couple of custom-based CNNs and TL-based fine-tuned CNNs have been implemented for segmenting the CT images of patients in the wake of the COVID-19 pandemic. These methods mainly extract the infected region and perform binary segmentation (COVID-19 infected or Normal lesion) or multi-class segmentation (GGO, consolidation, pleural effusion, crazy paving pattern, linear opacity, etc.) or both [29].

## 4.3 Multi-stage Techniques

A multi-stage technique performs classification as well as segmentation of COVID-19 infected regions.

There is no strict order for performing classification and segmentation in the DL models that use multi-stage techniques. However, the multi-stage framework reduced computational complexity and improves COVID-19 infection lesion diagnosis and analysis performance. Several studies claimed that employing a multi-stage framework can lead to better model diagnosis than single-stage diagnosis. In these studies, segmentation has been used as pre-processing for COVID-19 classification. Moreover, most of the studies performed segmentation after classification to accurately predict severity level and timely cure of COVID-19 confirmed cases.

## 4.4 Transfer Learning

TL leverages the knowledge learned by a pre-trained model and is fine-tuned for new challenges. However, building a DL model for new medical challenges from scratch is time-consuming and often gives less training accuracy because of limited data. Additionally, it consumes less computational power to fine-tune on a sufficient dataset compared to training from scratch[30]. Using a pre-trained model with few modifications or customization of additional CNN layers saves training time, leads faster convergence and improves model generalization [31], [32] because they already have an optimized learned pattern. The current pandemic has raised an urgency to find quick and efficient solutions to diagnose COVID-19 positive cases. Many pre-trained models are being used to classify and segment COVID-19 precisely and have proved equally beneficial. For COVID-19 analysis, different TL-based fine-tuned CNNs are shown in Fig. 5.

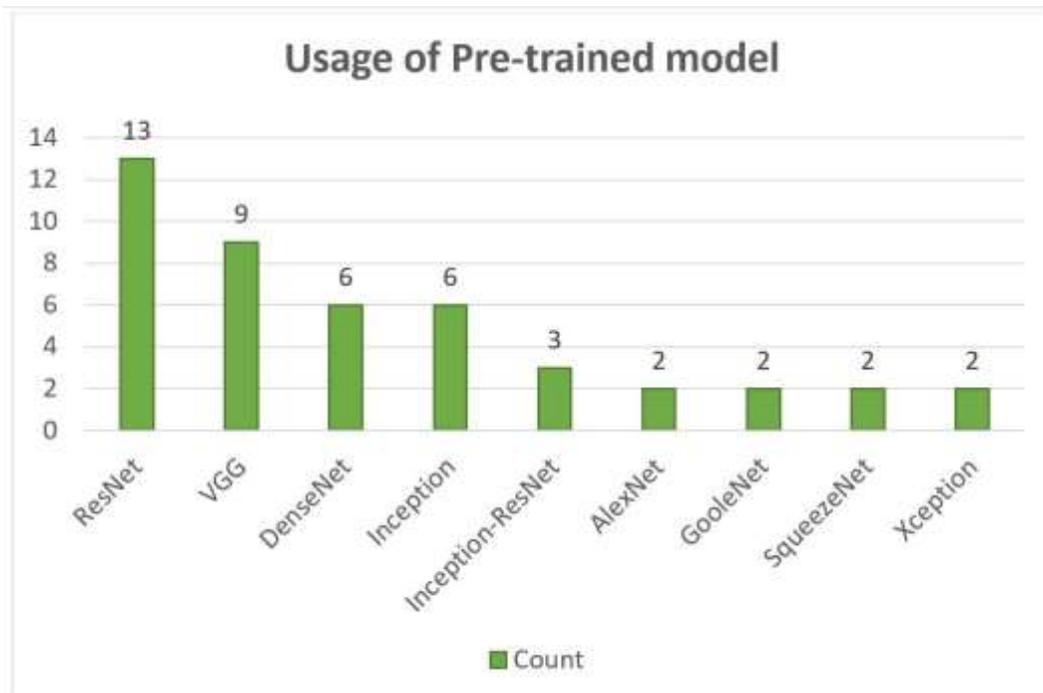

**Figure 5.** Usage of different TL-based pre-trained models for COVID-19 analysis

## 4.5 Custom-made Architecture

Custom-made deep CNNs are application-specific and can result in a consistent and better performing model [12]. These architectures evolve the existing DL techniques or design new models [33]. No weights and biases of the previous pre-trained model are used; however, this aspect increases computational cost and time [34]. The existing CNNs are modified by adding the initial and targeted layers to be compatible with the input dimensions and tuning them for radiological images. This can be achieved by hybridizing DL techniques (i.e., the fusion of homogenous DL techniques) [35], [36] or by hybridizing DL techniques with other areas in AI techniques [37], [38]. Developing custom DL methods for diagnosing COVID-19 infected patients is an active research area at the current time. These custom architectures are specifically developed to analyze COVID-19 affected lungs providing an efficient and accurate prediction of infected regions compared to the pre-trained model. The difference between a pre-trained model and a custom model is shown in Fig. 6.

## 5 Dataset and Performance Metrics

Despite numerous available researches on the COVID-19 pandemic, many researchers face the challenge of dataset collection. Privacy concerns led to limited public availability of COVID-19 related radiological datasets. These datasets contain COVID-19 infected samples, Pneumonia infected (viral or bacterial), SARS infected, and Normal persons. Most of these methods were developed using multiple source repositories, while some are based on single-source data. The accessed repositories comprised data from different accessible sources and hospitals, and radiologists verified these radiological images. In this survey, the performance of classification and segmentation models have been evaluated using standard performance metrics. The classification is assessed using accuracy (ACC), F1-score, sensitivity/recall (SEN), precision (PRE), specificity (SPEC), and area under the curve (AUC). In contrast, the segmentation is evaluated using Dice Similarity (DS) and the intersection of union (IoU). To ensure that the developed scheme generalizes well and selects appropriate hyper-parameters, different cross-validation (CV) techniques have been employed. Moreover, some additional classification and segmentation performance have been utilized for better evaluations. The performance of a multi-stage DL framework is normally assessed using the metrics mentioned earlier. The common performance measures used by most of the reported studies are as shown in Table 1.

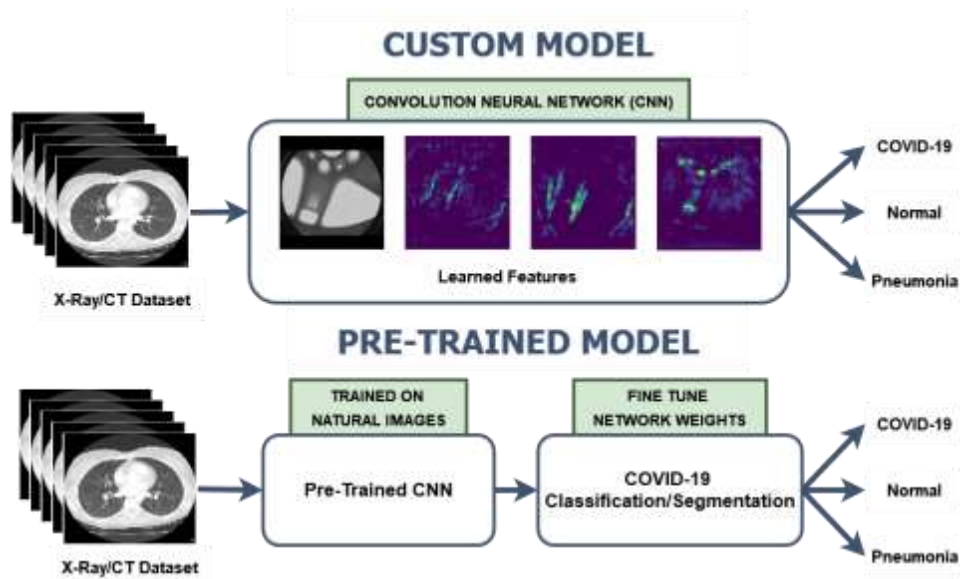

**Figure 6.** Difference between a custom model and a pre-trained deep CNN model

**Table 1.** Frequently used performance measures

| Classification | Segmentation | Formula |
|---|---|---|
| Accuracy (**ACC**) | Accuracy (**ACC**) | ACC = (TP+TN)/(TP+FP+TN+FN) |
| Precision (**PRE**) | Dice Similarity (**DS**) | PRE = TP/(TP+FP) |
| Specificity (**SPEC**) | Specificity (**SPEC**) | SPEC = TN/(TN+FP) |
| Sensitivity or Recall (**SEN**) | Sensitivity (**SEN**) | SEN = TP/(TP+FN) |
| F1-Score | Intersection over Union (**IOU**) | F1-score = 2*(PRE*SEN)/ (PRE + SEN) |
| Matthews Correlation Coefficient (**MCC**) | | DS= 2TP/(2TP+FN+FP) |
| | | IOU = TP/(TP+FP+FN) |
| where True-Positive (TP), True-Negative (TN), False-Positive (FP), and False-Negative (FN) | | |

# 6 Analysis of Existing Literature

A summary of the DL-based COVID-19 diagnosis studies related to classification, segmentation, and multi-stage techniques is shown in Fig. 7. The related literature reviewed against each technique is given in Table 2.

## 6.1 Analysis of Classification Techniques

### 6.1.1 Exploitation of TL in CNN

Largely, existing pre-trained CNNs have been employed and fine-tuned on radiological images using TL to classify COVID-19 positive cases precisely and have proved beneficial. A total of eleven DL-based pre-trained systems with deep TL have been reported so far. There are seven classification models employed for the CX-Ray and four classification models employed for the CT image analyses.

### Classification Using Chest X-Ray

Pham [39] studied the impact of fine-tuning three pre-trained CNN models using TL for COVID-19 classification: SqueezeNet, AlexNet and GoogleNet. The publically available dataset was collected from three authentic COVID-19 CX-Ray databases or Kaggle. The results by Pham [39] indicated that the fine-tuning of these pre-trained CNN models could lead to better classification performance without any pre-processing step like data augmentation. Additionally, it was found that these models required less training time than several other pre-trained models. The binary and ternary classification performance was evaluated using various ratios of testing and training sets, yielding six datasets. The three TL-based fine-tuned models showed the best performance metrics on two datasets for binary classification with 50% training and testing sets.

**Figure 7.** Papers reviewed for this article

**Table 2.** Technique, Image Modality and References

| Class | Image Modality | References |
|---|---|---|
| **Classification Techniques** | | |
| Binary | X-Ray | [39]–[47] |
| Binary | CT | [48]–[52] |
| Ternary | X-Ray | [39], [45], [46], [53]–[56] |
| Ternary | CT | [57] |
| Other | X-Ray | [46][58] |
| **Segmentation Techniques** | | |
| Binary | X-Ray | [59] |
| Binary | CT | [28], [29], [60]–[67] |
| Multi | CT | [28], [29], [60], [66], [68], [69] |
| **Multi-Stage Techniques** | | |
| Binary | X-Ray | [70] |
| Binary | CT | [71]–[73] |
| Ternary | X-Ray | |
| Ternary | CT | [71]–[73] |
| Other | X-Ray | [74] |

Narin et al. [40] reported an automated COVID-19 detection framework by employing five pre-trained existing CNN models and comparing classification performance for each model. These models are named as ResNet-50, ResNet-152, ResNet-101, Inception-ResNet-V2, and Inception-V3. The reported framework performed three different binary classifications (COVID-19 Vs. Normal, COVID-19 vs. viral Pneumonia, and COVID-19 vs. bacterial Pneumonia) using four different classes (bacterial Pneumonia, viral Pneumonia, COVID and normal). The ResNet-50 achieved the highest performance on all three different datasets, collected from Dr. Joseph's open-source GitHub repository [75], ChestX-ray8 database [76], and Kaggle repository named CX-Ray Images [77]. In the study by Asif et al. [53], a pre-trained inception-V3 DCNN model with TL performed automated COVID-19 detection via CX-Rays. The study aimed to diagnose and screen COVID-19 patients from Pneumonia and Normal CX-Rays. The dataset was imbalanced, comprised of three classes, having samples as follows: COVID-19 = 864, Pneumonia = 1345, and Normal = 1341 CX-Rays. The data was collected from various open-source like Dr. Joseph's GitHub repository [75], SIRM [78], COVID-19 Radiography Database [79]. The model was evaluated using accuracy and cross-entropy loss, achieved optimal results using TL and was easily deployable. Markis et al. [54] employed nine pre-trained existing DCNN models and fine-tuned on the ternary class dataset using TL to screen COVID-19 from Pneumonia and Normal CX-Rays. The samples were taken from two publicly available datasets; Dr. Joseph's GitHub repository [75] and Kaggle repository [77]. The evenly distributed dataset used for this study is collectively available at [80]. Abbas et al. [55] developed a new DCNN based DeTraC architecture to remove class irregularities through class decomposition, thus detecting COVID-19 CX-Rays from Normal and Non-COVID-19 patients. The classification was evaluated on a small dataset containing COVID-19 (105), Normal (80), and SARS (11) patients. The datasets were collected from the COVID-19 database [75] and JSRT [81], [82]. Khan and Aslam [40] employed four pre-trained DL classification models and fine-tuned them on a binary class dataset containing Normal and COVID-19 CX-Rays. The dataset of CX-Rays was collected from four different data sources: COVID-19 CX-Ray database by Cohen et al. [75], 25 samples of COVID-19 from data initiative [83], 180 from the COVID-19 CX-Ray [83]. Samples of both categories were taken (COVID-19 = 195, Normal = 862) from the COVID-19 radiography database [76]. In Islam et al. [42], a deep hybrid CNN and recurrent neural network named as CNN-RNN technique have been employed to classify COVID-19, Normal, and Pneumonia patients. Four pre-trained existing CNNs were used for deep feature extraction and then individually combined with RNN for ternary-class classification using CX-Rays. A balanced dataset of all adults was collected from seven different sources: repository of GitHub [75], [84], Radiopaedia [85], SIRM [78], data repository at Figshare websites [86], Kaggle [77], [87] and augmented images from Mendeley [88]. The experiments indicated that VGG19-RNN performed best concerning all the evaluation metrics and computational time. Additionally, the system employed Gradient-weighted Class Activation Mapping (CAM) for analyzing infection-specific regions to strengthen radiologist decisions. Table 3 provides the complete summary of the above discussion.

**Classification Using CT**

Jaiswal et al. [89] employed a TL-based pre-trained DenseNet-201 CNN on the CT dataset to screen COVID-19 infected samples from Non-COVID-19. The COVID-19 CT dataset was collected from Kaggle [90] consisting of 2492 samples. The performance of TL-based fine-tuned DesnseNet201 was analyzed and achieved significantly higher classification performance than other well-known pre-trained existing CNN models. A 2D DL classification technique, FCONet, was reported by Ko et al. [57] to discriminate between COVID-19 and Non-COVID-19 patients via CT samples. The FCONet technique included four pre-trained CNN models and fine-tuned CT scans. A total CT samples (3993) were collected from four sources. These include Chonnam National University Hospital, Wonkwang University Hospital, and SIRM database [78]. The test set consisted of CT samples gathered from some papers [94]. ResNet-50 outperformed among the other three CNN models when tested on low-quality CT images. Pathak et al. [49] reported a TL-based CNN framework to screen COVID-19 infection using a CT dataset. The framework implemented ResNet-50 to extract the potential deep features and custom-made CNN for the discrimination of COVID-19 and Non-COVID-19 patients. The developed system aims to handle the problem of bilateral change that occurs when using the CT image dataset. The model employed 10-fold cross-validation to prevent over-fitting and top-2 smooth loss function from overcoming noise and imbalanced dataset. The CT images dataset was gathered from several sources, including COVID-19 = 413 while Non-COVID-19 (Normal and Pneumonia=439) samples, but not limited to [95], [96]. According to Pathak et al., it is evident from the results that the developed model can replace COVID-19 testing kits. Moreover, Hasan et al. [50] performed classification between COVID-19 and Non-COVID-19 CT images using DenseNet-121 and achieved considerable performance of all classification metrics. The dataset for classification was obtained from a hospital in Sao Paulo, Brazil[87] containing COVID-19=1252 and Non-COVID-19=1230 samples. A summary of COVID-19 detection using standard performance metrics and the above discussion is given in Table 4.

**Table 3.** Classification: Pre-Trained DL Models with X-Ray Dataset

| Ref. | Data Sources | Technique | Classes | Performance |
|---|---|---|---|---|
| Pham [39] | Radiography Database[79] COVID-19 CX-Ray Dataset [83] COVID-19 CX-Ray Dataset Initiative [75] (50:50%) hold-out CV | AlexNet GoogLeNet **SqueezeNet** | Binary<br><br>Ternary | ACC=99.85%, F1-Score=99.8%, SEN=100%, PRE=99.57%, SPEC= 99.8%, AUC =99.9%<br>ACC=97.47%, F1-Score=96.3%, SEN= 98.48 %, PRE= 94.20%, SPEC= 97.39%, AUC =99.9% |
| Narin et al. [40] | Cohen repository [75] "ChestX-ray8" database [76]" CX-Ray Images (Pneumonia)" [77] | **ResNet-50**, ResNet-101 ResNet-152, Inception-V3 Inception-ResNet-V2 | Binary | **COVID-19 Vs Normal** ACC=96.1%, F1-Score= 83.5% SEN= 91.8%, PRE = 76.5% SPEC= 96.6% **COVID-19 Vs viral Pneumonia** ACC=99.5%, F1-Score=98.7% SEN=99.4%, PRE=98.0%, SPEC= 99.5% **COVID-19 Vs bacterial Pneumonia** ACC=99.7%, F1-Score= 98.5%, SEN= 98.8%, PRE = 98.3%, SPEC= 99.8% |
| Asif et al.[53] | GitHub repository [75] SIRM [78] COVID19 Database [79] | Inception V3 | Ternary | ACC = 98.30% (90:10%) Hold-out CV |
| Markis et al. [54] | GitHub repository [75] Kaggle's repository [77] GitHub repository [80] | **VGG16**, VGG19 MobileNetV2, InceptionV3 Xception, InceptionResNet-V2 DenseNet-201, ResNet152-V2 NASNetLarge | Ternary | ACC = 95.88%, F1-score=96.0%, PRE=96.0%, SEN=96.0%, SPEC=98.0% |
| Abbas et al.[55] 50 | JSRT [81], [82] CX-Ray samples [75] | **DeTraC** (Validated on AlexNet VGG19, ResNet GoogleNet, SqueezeNet) | Ternary | ACC = 97.35%, SEN= 98.23%, SPEC= 96.34%, ROC-AUC=96.5% |
| Khan & Aslam [41] | Cohen repository [75] COVID-19 CX-Ray [83] COVID-19 CX-Ray [91] COVID-19 database [79] | **VGG16**, ResNet50, VGG19, DenseNet121 | Binary | ACC = 99.33%, F1-Score = 99.28%, SEN= 99.28%, SPEC=99.38%, FNR= 72.0% FPR= 62.0%, PPV= 99.28% |

| Islam et al. [42] | GitHub repository [75], [84] SIRM [78], Radiopaedia [92], Figshare data repository [86][93] Kaggle [77], [87], Mendeley [88] | **VGG19-RNN** **(VGG19,** DenseNet121 InceptionV3 InceptionResNetV2) | Ternary | ACC = 99.85%, F1 = 99.78% SEN= 99.78%, PRE= 99.78, AUC = 99.99 % (80:20%) Hold-out CV |
|---|---|---|---|---|

### 6.1.2 Custom-made CNN-based Techniques

Largely, the existing DL models may not be effective for COVID-19 analysis because they are explicitly designed for natural scenes. Therefore, several techniques have been devised to capture the COVID-19 infection region, which is different from deformed and healthy regions. In this sub-section, custom-made DL classification models employed for the radiological image analysis have been reported.

**Table 4.** Classification: Pre-Trained DL Models with CT Dataset

| Reference | Data Sources and Cross-Validation | Technique | Classes | Performance |
|---|---|---|---|---|
| Jaiswal et al. [48] | SARS-CoV-2 CT on Kaggle [90] (85:15%) Hold-out CV | DenseNet201 | Binary | ACC = 96.3%, F1-Score = 96.3%, AUC-ROC = 97.0% |
| Ko et al. [57] | SIRM [78] CT Zhao [97] | FCONet (VGG16, **ResNet-50** Inception-v3, Xception) | Ternary | ACC = 99.87%, SEN = 99.6%, SPEC = 100% |
| Pathak et al. [49] | Various datasets [95], [96] (60:40%) 10-Fold CV | ResNet-50 based DCNN | Binary | ACC = 93.0%, SEN=91.5%, PRE= 95.2%, SPEC=94.8%, NPV=90.8% |
| Hasan et al.[50] | Hospital in Sao Paulo, Brazil [90] (70:30%) Hold-out CV | DenseNet-121 | Binary | ACC = 92.0%, F1-Score = 89%, SEN = 95.0%, PRE = 84.0%, GM= 89.0%. |

## Classification Using CX-Ray

A customized COVID-XNet CNN, proposed by Duran-Lopez et al. [43] has been employed to classify COVID-19 infected and Normal CX-Ray samples effectively. In this classification framework, the CX-Rays were enhanced using a set of pre-processing techniques such as histogram equalization and contrast enhancement and fed to the COVID-XNet (trained from scratch). The COVID-XNet architecture consists of five convolution layers, four max-pooling layers, and an average-pooling and softmax layer. The COVID-XNet employed on the imbalanced CX-Ray dataset consisted of 6926 samples (COVID-19=2589, Normal=4337) collected from Medical Imaging Databank of the Valencia Region (BIMCV) [98], the COVID-19 samples from Cohen et al. [75], and Padchest dataset by BIMCV [99]. Additionally, CAMs have been used for visualizing the infection, making COVID-XNet a promising diagnostic tool for COVID-19 infected cases. Gupta et al. found a new InstaCovNet-19 framework that performed binary (COVID-19, Non-COVID-19) and ternary classification (Pneumonia, Normal, COVID-19). This custom-made classification framework was built using a combination of pre-trained CNNs (InceptionV3, Xception, NASNet, ResNet101, and MobileNet) to counterbalance fewer data samples. The stacking layer combined the forecasts from the pre-trained existing CNNs where an integrated stacking unit at the target level has been employed to perform classification. The dataset consisted of CX-Rays and was collected from two publicly available repositories; (COVID-19 = 219, Pneumonia = 1345, Normal = 1341) from COVID-19 radiography database [43] and (COVID-19 = 142 samples to get total COVID-19= 361) from CX-Ray dataset [100]. Khan et al. [46] presented a customized deep CNN classification model named CoroNet that performs binary, ternary, and four-class classification (COVID-19, Normal), (Normal, Pneumonia, and COVID-19), and (COVID-19, viral and bacterial Pneumonia, Normal), respectively, using on CX-Ray dataset. The CoroNet architecture has been customized by utilizing and modifying pre-trained existing Xception. The CX-Ray dataset was gathered from two different publicly available databases; Joseph et al. open-source GitHub repository [75] and Kaggle [77]. A total of 1251 CX-Ray samples (Normal = 310, COVID-19 = 284, bacterial-Pneumonia = 330 and viral-Pneumonia = 327) were collected. Pereira et al. [58] performed multi and hierarchical-class classification between different types of Pneumonia, including the one which caused COVID-19. The COVID-19 classification framework was accomplished in four distinct phases: In phase 1, feature extraction using text descriptors (both handcrafted or static and non-handcrafted or deep (Inception-V3). Phase 2 and 3 performed early feature fusion (static + deep) and resampling, respectively. In the last phase, multi-class and hierarchical classifications using early fusion (Phase 1) and late fusion (Phase 2) have been performed. The dataset was collected from three sources which include Dr. Cohen's GitHub repository [75], Radiopaedia [76] and NIH dataset [22]. The dataset in multi-class consisted of 1144 samples (training set=802 and testing set=342). While the hierarchical class is made up of 1687 samples (training and testing set=1186, 501, respectively). Sahlol et al. [47] developed a classification technique by combining a pre-trained CNN (Inception) and swarm-based feature selection method (Fractional Order Marine Predators algorithm) to detect COVID-19 CX-Ray. The developed system was evaluated on two different datasets collected from various sources. Dataset1 contained 200 COVID-19 images gathered by Cohen, Morrison, and Dao [75] and 1675 Non-COVID-19 samples collected from CX-Ray Kaggle dataset [101]. Dataset2 was gathered from researchers at the University of Qatar, University of Dhaka, and fellows from Malaysia and Pakistan [96]. Some positive COVID-19 samples from the SIRM Database [102] were added in dataset2, consisting of (COVID-19=219, and Non-COVID-19=1341) images. Gour and Jain [56] proposed a stacked CNN model called CovNet30, which is built into a 30-layered model and used sub-models of VGG19. The proposed system performs classification between COVID-19 and Non-COVID-19 (Pneumonia and Normal) CX-Rays. A combination of modified datasets called COVID19CXr was generated using three publicly available sources [75], [83], [103]. COVID19CXr consisted of a total of 2764 CX-Rays (COVID-19 = 270, Pneumonia = 1355, Normal = 1139). The technique validated two datasets and outperformed comparative analysis with several existing models. A summary of the above discussion is given in Table 5.

**Classification Using CT**

Wang et al. [51] reported a two-fold customized DL-based technique named FGCNet for COVID-19 detection. FGCNet extracts individual image-level representations using a convolution neural network in the first fold. While in the second fold, a convolutional graph network is incorporated in FGCNet to capture the relation-aware representations from the COVID-19 infected images. The classification was performed between 320 COVID-19 CT samples and 320 healthy control (HC) CT images and evaluated using standard metrics. The additional performance metrics include Matthew correlation coefficient (MCC) and Fowlkes–Mallows index (FMI).

**Table 5.** Classification: Custom DL Models with X-Ray Dataset

| Ref. | Data Sources and Cross-Validation | Technique | Classes | Performance |
|------|-----------------------------------|-----------|---------|-------------|
| Duran-Lopez et al [43] | BIMCV [98]$^{91}$, Cohen repository [75]$^{74}$ Padchest dataset [99]$^{92}$ (80:20%) 5-Fold CV | COVID-XNet (Custom-made CNN) | Binary | ACC = 94.43%, F1 = 93.14%, SEN=92.53%, SPEC = 96.33%, AUC-ROC = 98.8% |
| Gupta et al [45] | COVID-19 Radiography Database [45] CX-Ray dataset [104] (80:20%) Hold-Out CV | InstaCovNet-19 (Integrated Deep CNN) | Binary Ternary | ACC = 99.5%, SEN = 99.0%, PRE = 100% ACC=99.1%, PRE = 99.0%, SEN= 99.0%, F1-score= 99.0%, |
| Khan et al. [46] | Cohen repository [75] "CX-Ray Images (Pneumonia)" [77] (80:20%) 4-Fold CV | CoroNet CNN | Binary Ternary Multi-4 | ACC=99.0%, F1-Score=98.5% PRE=98.3%, SEN=99.3%, SPEC=98.6% ACC=95.0%, F1-Score=95.6% PRE=95%, SEN=96.9%, SPEC=97.5% ACC=89.6%, F1-Score=89.8%, SEN=89.92%, PRE=90.0%, SPEC=96.4% |
| Pereira et al. [58] | Cohen repository [75], Radiopaedia [76] NIH dataset [22] (70:30%) Hold-Out CV | Pre-trained CNN Feature-ML classifiers | Multi Hierarchical | F1 = 65.0% F1 = 89.0 % |
| Sahlol et al. [47] | Cohen repository [75] Kaggle CX-Ray [101] Team of researchers [96] SIRM [102] | Inception based FO-MPA CNN | Binary | dataset1: ACC= 98.77%, F1-Score == 98.21 % dataset2: ACC= 99.68%, F1-Score= 99.08%, (80:20%) 5-Fold CV |
| Gour & Jain [56] | COVID19CXr, generated using [75], [83], [103](80:20%) 5-Fold CV | CovNet30 | Ternary | ACC = 92.7%, F1-Score= 93.0%, PPV = 92.1%, SEN= 93.3%, SPEC= 95.8%, AUC-ROC = 95.0% |

In another study by Singh et al. [95], a multi-objective differential evolution-based CNN, MODE, performs classification between COVID-19 and negative CT samples. Classification between COVID-19 CT images has been first performed using CNN, ANN, and ANFIS. Then, the MODE algorithm has been used to fine-tune CNN and optimize hyper-parameters such as learning rate, kernel size, activation function, etc. The chest CT images were taken from [105], which included 73 COVID-19 infected patients. Several experimentations with various training and testing ratios were taken into account respectively (20:80%, 30:70%, 40:60%, 50:50%, 60:40%,70:30%, 80:20%, and 90:10%). A summary of the above discussion is given in Table 6.

## 6.2 Analysis of Segmentation Techniques

### 6.2.1 Exploitation of TL in CNN

In this analysis, a limited amount of reported studies is available to segment COVID-19 infected radiological images by exploiting the concept of TL in existing CNN models. However, a few pre-trained CNN models have been utilized and fine-tuned on CT scans and X-Ray images using TL. Two and one segmentation techniques are employed for the CT and CX-Ray image analysis, respectively. A summary of these techniques is given in Table 7.

**Segmentation Using X-Ray**

In a study by Tang et al. [106] U-Net architecture was applied as a segmentation model for X-Ray images with ResNet-18 backbone. As a preprocessing step all images were resized to same size and cropped to show only lung regions. To increase the size of dataset, data was augmented via several methods. The dataset [107] developed from 221 CX-Rays of COVID-19 patients from rural areas of United States.

**Table 6.** Classification: Custom DL Models with CT Dataset

| Ref. | Data Sources and Cross-Validation | Technique | Classes | Performance |
|------|-----------------------------------|-----------|---------|-------------|
| Wang et al.[51] 46 | Dataset acquired using Philips Ingenuity 64 row spiral CT machine (90:10%) Hold-out CV | **FGCNet,** RBFNN [104] RCBBO [108] ,COVNet [109] | Binary | ACC = 97.14%, F1-score = 97.2%, MCC = 94.3%, SEN = 97.7%, PRE= 96.6%, SPEC = 96.6%, FMI = 97.2%. |
| Singh et al. [95] | Dataset [105] Kaggle CXR data [77] | **MODE-based CNN Model** ANN, ANFIS | Binary | ACC = 97.9%, F1-Score = 98.0% (20:80%) 10-Fold CV |

Another dataset [110] was also used from which 25 CX-Ray images were extracted to be used as validation set. They also showed that these predicted percentage of opacity region can prove helpful in predicting severity of the COVID-19 patients.

**Segmentation Using CT**

A study by Li et al. [68], developed an automatic system to quantify COVID-19 infected chest areas on thick CT sections. A deep CNN-based system applied U-Net's model and had Resnet-34 as the backbone to predict multiclass lesion segmentation. Moreover, a 2D-UNet was also trained to perform the lung segmentation. The CT scans of patients were collected from The First Hospital of Changsha. The dataset was divided into groups by severity of COVID-19 infection containing 79 CT scans of severely infected patients, 452 CT scans of generally infected patients, and 538 CT scans of Pneumonia infected patients gathered via multiple medical centers. The testing set consisted of 30 CT scans of COVID-19 patients in which 3 cases were severe, and 27 cases were non-severe. The portion of infection (POI) and the average infection Infection Hounsfield unit (iHU) were used to distinguish the severity level of COVID-19 infection.

Saood et al. [60] compared the performance of SegNet and UNet to predict the binary and multi-class COVID- 19 infected regions segmentation using CT Lungs samples. The SegNet is a scene segmentation network comprising encoder-decoder architecture, while UNet is a medical image segmentation network comprising encoders and expansive decoder with depth concatenations and skip connection paths. Both these models are trained and evaluated using a dataset collected by SIRM [110]. The training was performed by using nine different experimental settings via varying batch sizes and learning rates for both models and evaluated using standard segmentation metrics. The additional segmentation metrics include Global Mean (GM) and F-score. The summary is reported in Table 7.

**Table 7.** Segmentation: Pre-Trained DL Models with CX-Ray and CT Dataset

| Reference | Data Sources and Cross-Validation | Technique | Classes | Performance |
|---|---|---|---|---|
| **X-Ray Dataset** | | | | |
| Tang et al. [106] | Rural areas of US[107] Dataset [110] | U-Net withResnet-18 backbone | Binary | ROC-AUC=82.0%, IOU = 73.8% (70:30%) Holdout CV |
| **CT Datasets** | | | | |
| Li et al. [68] | The First Hospital of Changsha Multicenter CAP CT dataset | 2D UNet with Resnet-34 backbone | Multi-class | Lungs Infection: DS=74.0%, Severe vs Non-Severe AUC=97.5%, kappa Score=82.2% POI: SEN=92.41%, SPEC=90.49%, AUC=97.0% iHU: SEN= 91.14%, SPEC= 41.59%, AUC= 69.0% POI + iHU: SEN= 93.67%, SPEC= 88.05%, AUC= 97.0% |
| Saood et al. [60] | COVID-19 CT Segmentation dataset [110] (90:10%) 5-fold (CV) | **SegNet** UNet | Binary / Multi-class | ACC = 95.4% SEN=95.6%, SPEC= 95.4% , DS=74.%9, GM=95.5%, F-Score=86.1% ACC = 90.4%, SEN=67.33, SPEC=95.9%, DS=31.7%, GM=79.8%, F-Score=42.9% |

### 6.2.2 Custom Architectures based Techniques

A total of 16 custom DL-based techniques were reviewed. All these methods models used CT scan datasets.

**Deep Learning-based Infection Segmentation Using CX-Rays**

Currently, a minimal amount of literature is available that solely performed segmentation on CX-Ray using a custom-made CNN. Even though X-Rays are cheap and more easily accessible, performing segmentation on CX-Ray is challenging because of the projection of the ribs onto soft tissues in 2D images, making it difficult to locate the infected region of the lungs [111]. Additionally, there is a lack of annotated COVID-19 lung infection labeled data resulting in no custom-made CNN segmentation model for CX-Rays. Some available studies performed both segmentation and classification using X-Ray, which is discussed in section 5.

**Deep Learning based Infection Segmentation Using CT Lungs Images**

A customized DL model, Inf-Net, developed by D. Fan et al. [28] performed binary (Infection and Background) and multi-class (GGO, Consolidation, and Background) segmentation of infected region using CT scan samples. The developed model comprises five convolution layers, where the first two layers have been used to extract low-level features and passed to Edge Attention Module to enhance the edge information. These features have been passed to the next three layers to extract high-level features, fed in parallel to a partial decoder to generate maps globally and into several Reverse Attention modules to extract infected regions. To overcome the shortage of labeled data, they introduced Semi-Inf-Net that progressively generates labels for unlabeled data by enlarging the limited number of labeled training samples and generating the COVID-19 Semi-Seg dataset. This semi Inf-Net has been further extended to be used FCN8s or U-Net model as a backbone for multi-class labeling of lung infection. The COVID-SemiSeg dataset is based on total of 1710 samples generated from COVID-19 CT Segmentation[106] (COVID-19 = 110 labeled) and COVID-19 CT Collection [75] (1600 unlabeled images). The performance evaluation parameters also included Structure Measure (Sα), Enhance-alignment Measure (EAM), and Mean Absolute Error (MAE).

Wang et al. [61], developed a model for binary challenge using COVID-19 Lesion segmentation network (COPLE-Net) based on U-Net. The noise-robust model utilized dual pooling (max-pooling + average pooling) and was used as a bridge between the encoder and decoder to reduce the gap between the low- and high-level attributes. Additionally, a module named atrous spatial pyramid-pooling was employed to perform better lesion segmentation at multiple scales. They also introduced a novel loss function, LNR-Dice, robust against noisy labels. COPLE-Net and LNR-Dice were integrated into a self-ensemble way to overcome the challenge of noisy training labels. An exponential moving average and aggregation of a framework as backbone were utilized in this framework to learn a target level model. The dataset contained 558 COVID-19 infected patients, which were retrieved from ten hospitals and captured from various CT scan machines.

A study by Zhang et al. [61], developed a conditional generative model named CoSinGAN. The model steadily generated high-resolution and realistic radiological images from a single real CT slice. It consisted of three components: a multi-scale two-staged pyramidal generative adversarial networks (GANs) framework, which dealt with the increase in image resolution, compatible with the input condition, and restoring image details. Second, they used a mixed reconstruction loss by combining four losses, which give rich and stable gradient information. Thirdly, a hierarchical data-augmentation based on robust and weak-augmentation has been used to learn conditional distribution from the image. Moreover, the variants of CoSinGANs: IF-CoSinGAN (image fusion) and RC-CoSinGAN (randomizing input conditions), have been developed to generate diverse images. The CT image dataset was gathered from open sources; 20 CT scans from COVID-19-CT-Seg 106, 50 MosMed CT samples 47. The quality of radiological images generated by CoSinGAN was tested by employing the U-Net model on these images and using an additional metric called normalized surface dice (NSD) for segmentation evaluation.

In another study, Yao et al. [63] developed a normalcy-converting network (NormNet), a label-free segmentation method to differentiate the COVID-19 infection from healthy region. The operations like random shape, salt noise, and variation intensity have been employed on healthy CT datasets to generate synthetic lesions to construct the normal and abnormal image dataset for the robust training. NormNet with 3D U-Net backbone has been trained on the robust dataset using ensemble learning of five models, and prediction has been taken based on majority-vote for healthy regions. Moreover, the technique performed data normalization, padding, and centralization as a pre-processing step to obtain the lesion segmentation. Moreover, additional pre-processing steps included subtracting the predicted healthy region from the lung mask, mean filtering, and region growing algorithm improved the infection segmentation performance. Multiple datasets used in this model were LUNA16 [112] as CT data for healthy lung modeling, Coronacases [113] and Radiopaedia [85] for COVID-19 infection segmentation, and NSCLC left and right lung segmentation [114], [115], StructSeg lung organ segmentation [116], MSD Lung tumor segmentation [117] as CT data for general-purpose lung segmentation. The training dataset has been collected from LUNA16, while the testing is from two datasets: Coronacases and Radiopaedia. The developed model has been compared with various already existing DL techniques on the lung regions dataset to access the lung segmentation capability.

Zhou et al. [64] designed a COVID-19 binary segmentation and quantification framework using CT Lung images. The developed framework introduced a pre-processing method that normalized the CT samples' resolution, dimension, and pixel intensities. Moreover, the pre-processing steps involved spatial and signal normalization and data augmentation. Augmentation was done via modeling and simulating dynamic infection changes to deal with the limited labeled data. A Three-Way segmentation model has been developed for segmentation, which divides the 3-dimensional segmentation issue into three 2-dimensional by employing three independent 2D U-nets. The COVID-19 infected patient datasets contained CT scans gathered from five hospitals of two countries. These datasets were named as Harbin dataset (201 CT scans) and Riyadh dataset (21 CT scans). Five-fold CV was performed on samples collected from Harbin (one-fold reserved for testing while the remaining four folds were employed for training and validation).

Mishra [118] developed a modified U-Net framework that performed binary segmentation for lesion-infected lung regions. In the pre-processing stage, contrast limited adaptive histogram equalization technique has been used to enhance the image contrast. The U-Net has been modified by incorporating batch normalization and modified transposed convolution layers. Zenodo dataset [119] which contained CT samples of COVID-19 victims, was used with different validation splits like three, four, and seven-fold CV methods. The additional performance evaluation metrics included precision and recall.

A VB-Net segmentation model reported by Shan et al. [69] integrated V-Net with a bottleneck model. The bottleneck sub-network structure included three convolutional layers in contrast to V-Net, which utilized a simple convolution operation at the bottleneck. The VB-Net has been trained via the Human-Involved-Model-Iterations technique. They collected a custom dataset from different hospitals from Shanghai and outside Shanghai, consisting of 249 sample data of CT scans for training and 300 validation sample CT scans data. A comparative analysis with U-Net revealed that the developed model performed best with 0.3% of POI Error of the whole lung.

Chassagnon et al.[65] investigated and developed an AI-based strategy for quantifying and staging COVID-19 CT scans. The introduced method named COVID-ENet performed binary segmentation and was designed by ensembling 2 and 3-dimensional CNNs. The COVID-E2D architecture was based on AtlasNet 2D architecture, and ML classifiers which were trained samples, were registered from predefined templates. COVID-E3D architecture is based on 3D-UNet and performed automatic multi-omics COVID-19 profiling and staging with the help of lung segmentation. The cardiac and other disease information has been extracted from images by integrating biological and clinical measures. Moreover, the staging and prognosis have been estimated using an ensemble learning of various ML algorithms. The dataset of 693 patients is formed using the data provided by eight large University Hospitals. For COVID-19 infection segmentation, the training and validation set contained data of 50 patients, while the testing set contained data of 130 patients. While for the staging and prognosis, the training and validation set contained data of 536 patients, and data of 157 patients were used to form a testing set. To evaluate the quality of automated segmentation, the developed model was performed with DS, STD, and a mean score of Hausdorff distance metrics. Various results for COVID-19 staging and prognosis have also been reported.

A study by Ni et al. [66] introduced an automatic detection and evaluation system to detect patients with COVID-19 Pneumonia. Architecture performed three tasks: 1) MVP-Net utilized for automatic lesion detection, 2) 3D U-Net was exploited to classify voxels, and 3) 3D U-Net-based segmentation network was adopted to provide the pulmonary lobe segmentation. The dataset used in this study was taken from seven different hospitals located in China and consisted of 19,291 CT samples (COVID-19 = 3854, Pneumonia = 6871, healthy = 8566). The technique has been tested independently where the testing set contained 96 sample images of COVID-19 patients. The diagnostic performance of the customized model versus radiology residents was employed for patient-level and Lung lobe level segmentation.

In another study Chen et al., [67] used a UNet++ architecture to detect viral Pneumonia in CT scans trained for binary lesion lung segmentation. The dataset of CT images was gathered from Wuhan University's Renmin Hospital, China. A total of 35,355 samples were collected from 106 patients admitted to hospitals. Among the dataset, 20886 images were gathered from 51 patients with COVID-19 Pneumonia and 14,469 images from Non-COVID-19 patients. The training set contained 691 COVID-19 Pneumonia lesion samples and 300 control patients' samples. The testing retrospective dataset contained 636 samples of COVID-19 lesions and 9369 CT scans of control patients, while the prospective testing dataset contained 8172 and 5739 for COVID-19 lesions and control patients CT scans, respectively. Performance evaluation parameters additionally included NPV and PPV. The website implementation is an open-access platform to help diagnose COVID-19 via uploading patient CT samples.

Tilborghs et al. [29] conducted a benchmark study of twelve DL frameworks for automated lung and infections segmentation in CT images of the COVID-19 patients. The standard dataset was collected from different medical centers in Latin America and Europe, containing 52 COVID-19, 14 Non-COVID-19, and 7 CT scans of suspected COVID-19 patients. Largely, these methods used the different settings of re-sampling, clipping HU intensities, and up-sampling as a pre-processing for automatic lung segmentation of CT scans.   The performance was also evaluated using additional metrics like average-surface-distance (ASD), HD at percentile 95% (HD95), and absolute-volume-difference (AVD) for binary and multiclass lesion segmentation.

The JoHof [120], DMLu, and DMLo were used to automatically segment lungs and lobes. Segmentation models for COVID-19 like InfNet [28], COVID-ENet [65] and CTA [67] are already discussed in this study. 2DS is a 2D U-Net comprised of the convolutional block, ReLU activation, dropout, max-pooling, and transposed convolutions and trained using a patch-based strategy to perform binary lesion segmentation. In addition to the standard data, 50 CT scans from publically available MosMedData[47] were also used for training. Moreover, the model DMmc, which was based on DeepMedic used in this study, performed multiclass lesion segmentation. Furthermore, the output layer of the DeepMedic network is changed with three outputs for the segmentation of each of the classes: GGO, CON, and CPP. Finally, a method named 2DRnx was developed by using a 2D U-Net to perform 2D lesion classification for multiclass segmentation.

WASS performed binary lesion and multiclass lesion lung segmentation via a 3-dimensional U-Net-based framework. The framework used a loss function called Generalized Wasserstein Dice to train the DL model by benefiting its hierarchal structure. A modeling framework known as distributionally robust optimization has been used to apply hardness weighted sampling for training the model to tackle the problem of the high variability of pathologies. Moreover, data augmentation has been used as a pre-processing step for increasing robustness and improving generalization.

UNWM method performed lung, binary and multi-class lesions segmentation using waterfall masking. The model was comprised of two main parts: 1) A 3D-based U-Net with modified kernels and convolution layers, 2) waterfall masking technique employed for segmentation of class (lung, binary and multi-class segmentation). The waterfall masking consisted of two independent pathways, resulting in six parallel and independent pathways. Different data augmentation techniques like Gaussian noise, translation, rotation, and flipping were applied to reduce the overfitting. The last model named MAJ performed ensemble learning by taking a vote from each of the aforementioned techniques. The segmentation for each ternary class was predicted by considering the majority vote. The result showed that the majority voting technique achieved the highest performance compared to existing models. Moreover, it can be concluded that no single model achieved significantly great results for all evaluation metrics compared to other models. Each model outperformed in different areas by employing individually; however, overall good performance was achieved through ensemble learning. A summary of the above discussion is given in Table 8.

## 6.3 Analysis of Multi-Stage Techniques

### 6.3.1 Exploitation of TL in CNN

In this section, a study on a multi-stage framework employing classification and segmentation techniques on radiological datasets, has been reported for COVID-19 infection analysis.

**Table 8.** Segmentation: Custom DL Models with CT Dataset

| Reference | Data Sources and Cross-Validation | Technique | Classes | Performance |
|---|---|---|---|---|
| Fan et al. [28] | COVID-19 CT Segmentation data [121]<br>COVID-19 CT Collection [75]<br>(75:25%) Hold-out CV | Inf-Net, Semi-Inf-Net, Semi-Inf-Net & MC, Semi-Inf-Net & FCN-8 | Binary<br><br>Multi-class | Inf-Net: DS = 68.2%, SEN = 69.2%, SPEC= 94.3%, S$\alpha$ = 78.1%, **Emean** = 83.8%, MAE = 8.20% (Binary)<br>Semi-Inf-Net: DS = 73.9%, SEN = 72.5%, SPEC= 96.0%, S$\alpha$ = 80.0%, **Emean** = 89.4%, MAE = 6.40% (Binary)<br>Semi-Inf-Net & FCN8s: DS = 47.4%, SEN = 47.8%, SPEC = 87.5%, S$\alpha$= 64.1%, **Emean** = 72.3%, MAE = 5.80% (Multi-class)<br>Semi-Inf-Net & MC: DS = 54.1%, SEN = 56.4%, SPEC = 96.7%, S$\alpha$= 65.5%, **Emean** = 82.8%, MAE = 5.70% (Multi-class) |
| Wang et al. [61] | 10 different hospitals | COPLE-Net based on UNet | Binary | DS = 80.29%, RVE =17.72%, HD95 = 18.72mm. |
| Zhang et al. [62] | COVID-19 CT Segmentation dataset [121]<br>MosMed dataset [47]<br>(80:20%) 5-fold CV | CoSinGAN, IF-CoSinGAN<br>**RC-CoSinGAN** | Binary | DS= 64.1%, NSD= 60.5%<br>DS=64.7%, NSD= 61.4%<br>DS=71.3%, NSD=72.0% |
| Yao et al. [63] | LUNA16 [112] Coronacase [113] Radiopaedia [85] NSCLC lung segmentation[114], [115] StructSeg lung organ segmentation [116], MSD Lung tumor segmentation [117] | NormNet based on 3D UNet | Binary | Coronacase: ACC = 87.8%, SPEC = 80.2%, SEN =78.8%, PREC = 90.6% Radiopaedia: ACC = 89.5%, SPEC = 87.9%, SEN =70.7%, PREC = 93.5% (80:20%) 5-fold CV |
| Zhou et al. [64] | Harbin dataset Riyadh dataset | Custom model based on 2D UNet | Binary | SEN = 77.60%, DS = 78.30% (80:20%) 5-Fold CV |
| Mishra [118] | Zenodo [119] (70:30%) 7-fold CV | Modified U-Net | Binary | DS = 96.91%, PRE= 96.42% SEN= 97.4%. |
| Shan et al. [69] | Custom dataset collected from hospitals (90:10%) 10-Fold CV | VB-Net based on VNet | Multi-class | DS = 91.6% POI estimation error=0.3% |
| Chassagnon et al. [65] | Custom dataset collected from 8 hospitals | COVIDENet | Binary | DS = 70.0%, STD= 12%, HD=8.96mm 80:20% Hold-out CV |
| Ni et al. [66] | Custom dataset collected from 7 hospitals of China (80:20%) Hold-Out CV | Custom model based on MVP-Net & 3D UNet | Binary Multi-class | Patient Level Segmentation: ACC = 94.0%, F1 = 97.0%, SEN = 100%, PPV = 94.0%, NPV = 100%, SPEC = 25.0% Lung lobe level segmentation ACC = 82.0%, SEN = 96.0%, SPEC = 63.0%, PPV = 78.0%, NPV = 93.0%, F1 score =86.0% |
| Chen et al. [67] | Custom dataset collected from Renmin Hospital, China | Custom model based on UNet++ | Binary | Internal (Retrospective Testing) Patient level: ACC = 95.2%, SEN = 100%, SPEC = 93.6%, PPV = 84.6%, NPV = 100%. Image ACC = 98.9%, SEN = 94.3%, SPEC = 99.2%, PPV = 88.37%, NPV = 99.6% Patient:(Prospective Testing) ACC = 92.6%, SEN = 100%, SPEC = 81.8%, PPV = 88. 9%, NPV = 100% External Patient:( Retrospective Testing) ACC = 96.0 %, SEN = 98.0%, SPEC = 94.0%, PPV = 94.2%, NPV = 97.9% |
| Tilborghs et al. [29] | Custom dataset from Europe & Latin America MosMed dataset [52] | 2DS based on 2D UNet (80:20%) 5-fold CV | Binary | DS = 63.9%, HD95 =100, ASD = 24.6 and AVD=132 |
| Tilborghs et al. [29] | Custom dataset from Europe & Latin America | DMmc based on DeepMedic network | Multi-class | DS= 46.0%, HD95 = 184 mm, ASD =123 mm, AVD = 93.6 mm |
| Tilborghs et al. [29] | Custom dataset from Europe & Latin America | 2DRnx based on 2D UNet | Multi-class | DS = 38.3%, HD95 =167mm, ASD = 107 mm, AVD = 149 mm |
| Tilborghs et al. [29] | Custom dataset from Europe & Latin America | WASS based on 3D UNet | Binary Multi-class | DS = 60.0%, HD95 = 77.6 mm, ASD = 25.3 mm, AVD = 176 mm DS = 20.0% , HD95 = 163 mm, ASD = 103 mm, AVD = 238 mm |
| Tilborghs et al. [29] | Custom dataset from Europe & Latin America | UNWM | Binary Multi-class | DS = 60.0%, DS = 64.9%, HD95=102 mm, ASD = 46.4 mm, AVD = 94.2 mm DS= 50.0% , DS = 46.5%, HD95=208 mm, ASD = 142 mm, AVD = 222 mm |
| Tilborghs et al. [29] | Custom dataset from Europe & Latin America | MAJ | Binary Multi-class | DS = 70.0%, DS = 72.4% , HD95=83.6 mm, ASD = 37.0 mm, AVD = 91.3 mm DS = 40.0%, DS =46.9%, HD95=166 mm, ASD = 109 mm, AVD = 152 mm |

## Multi-Stage Techniques Using X-Ray

Goldstein et al. [70] detected COVID-19 positive samples from Non-COVID-19 using a pre-trained ResNet50. The reported model enhanced classification performance by employing data augmentation and lung segmentation as pre-processing. The nearest neighbor algorithm was also applied as a classifier by replacing the softmax layer to detect COVID-19 CX-Ray effectively. The imbalance dataset was collected from four hospitals in Israel containing 1384 patients' CX-Rays (COVID-19 samples= 360 while Non-COVID-19 = 1024).

Kikkisetti et al. [74] employed a pre-trained VGG16 and fine-tuned TL to screen COVID-19

infections in CX-Rays. The multi-class classification was performed between COVID-19, Normal, and Pneumonia CX-Rays. The number of CX-rays collected samples for each class were: COVID-19=455, Normal=532, Pneumonia (bacterial) = 492, and Pneumonia (viral) = 552. The samples were collected from Cohen, Morrison, and Dao [75] and Kaggle sources [77]. This study employed a custom-made twenty-two convolutional layers CNN for lung infection segmentation and provided for multi-class classification. The segmented lungs' classification performance was better than the whole portable CX-Ray. A summary of the above discussion is given in Table 9.

**Multi-Stage Techniques Using CT**

Wang et al. [71] classified COVID-19 samples from other Pneumonia by performing diagnosis and prognosis using an automated DL system. This framework accomplished three tasks using two DL networks. In this regard, lung-segmentation and suppressing non-lung regions were performed using DenseNet121-FPN. In contrast, COVID-19Net performed diagnosis and prognosis. The multi-regional dataset was collected from 7 different hospitals in China. 4106 CT scans of people with lung cancer with EGFR gene sequencing were used to train the model. The study used an auxiliary training dataset with 4106 lung cancer patients with EGFR gene sequencing. The training set consisted of 709 patient samples (COVID-19 samples = 560, Pneumonia (other) = 149). Four external validation sets were used in this study with 226, 161, 53, and 117 CT samples, respectively. The diagnostic evaluation of the model includes additional metrics like calibration curves and the Hosmer-Lemeshow test. While the prognostic analysis was performed using the log-rank test and Kaplan-Meier. A summary of the above discussion is given in Table 10.

### 6.3.2 Custom Architectures based Techniques

For this section, six relevant papers were reviewed. Among these papers, 3 studies performed classification and segmentation using pre-trained models, while 6 studies employed custom DL techniques to accomplish the task of classification and segmentation.

**Table 9.** Integrated Technique: Pre-Trained Deep CNN Models using CX-Ray Images

| Reference | Data Sources and Cross-Validation | Technique | Classes | Performance |
|-----------|-----------------------------------|-----------|---------|-------------|
| Goldstein et al. [70] | Four hospitals of Israel Hold-out CV (85/15%) | ResNet50 | Binary | ACC = 89.7%, AUC = 0.95<br>SEN= 87.1%, SPEC= 92.4%, AUC-ROC= 95.0% |
| Kikkisetti et al. [74] | COVID-19 portable CXR [75]<br>Kaggle CXR data [77]<br><br>5-fold CV (75/25%) | Classification: VGG16<br>Segmentation: Custom CNN | 4-class | **Segmented Lungs'**<br>Classification: ACC=88.0%, SEN=91.0%, SPEC= 93.0%, AUC-ROC=89.0%<br>Segmentation: IoU=95.6%, DS = 97.2%<br>**Whole CX-Ray**<br>ACC=79.0%, SEN=79.0%, SPEC=93.0%, AUC-ROC=85.0% |

**Table 10.** Integrated Technique: Pre-Trained Deep CNN Models using CT Lung Images

| Reference | Data Sources | Technique | Classes | Performance |
|-----------|--------------|-----------|---------|-------------|
| Wang et al. [71] | Multi- regional Dataset from 7 different hospitals of China | Segmentation: DenseNet121-FPN Classification: COVID-19Net | Binary | ACC = 81.24 %, F1-score = 86.92%, SEN= 78.93%, SPEC=89.93% |

**Table 11.** Integrated Techniques: Custom DL Models with X-Ray

| Reference | Data Sources | Technique | Classes | Performance |
|-----------|--------------|-----------|---------|-------------|
| Aslan et al. 116 | COVID-19 radiography database [96] 5-fold CV | **vmAlexNet + BiLSTM** mAlexNet | Ternary | ACC=98.7%, PRE=98.8%, AUC-ROC=99.0%, Recall=98.8%, SPEC=99.3%, F1-score =98.8%, MCC=98.1%, kappa =97.07% |

**Multi-Stage Techniques for COVID-19 Diagnosis Using CX-Ray**

In a study by Aslan et al. [119], a two-stage DL architecture performed lung segmentation and ternary classification using an ANN and hybrid CNN model. The classification model contained Bidirectional Long Short-Term Memories (BiLSTM) layer and used AlexNet as the baseline model. The radiography database contains 2905 samples collected from different sources with 219 COVID-19, 1345 viral Pneumonia, and 1341 samples of Normal CX-Rays. A summary of the above discussion is given in Table 11.

**Multi-Stage Techniques Using CT**

Amyar et al.[117] developed a Multi-task learning deep CNN that performed three tasks, i.e., classification (COVID-19, Normal, other infections), lesion segmentation, and image reconstruction. The multi-task learning framework has been implemented as an encoder-decoder block. A CT image is input to an encoder, and output has been fed into three decoders used to accomplish three distinct tasks, i.e., image reconstruction, classification and then segmentation. The model used U-Net as the base model with a few modifications such as increasing the number of filters, replacing pooling with the strided-convolution to maintain spatial information, etc. The dataset was taken from three different sources: (lung cancer =98, Normal= 425) were collected from a cancer center in France named Henri Becquerel Cancer Center. The other two open-source datasets are taken from a GitHub repository; (COVID-19 = 347, Non-COVID = 397), and (COVID-19 = 100).

Wang et al.[118], used a prior-attention residual-learning strategy constituted PARL blocks to perform multi-task, lobe segmentation via 3D-UNet, and classification via 3D-ResNets. The classification has been performed between COVID-19, ILD (Interstitial Lung Diseases), and Non-Pneumonia samples. In contrast, Lobe segmentation has been performed as a pre-requisite for classification challenges. A total of 4657 CT scans were collected from different hospitals, among 936 included Normal, 2406 ILD, and 1315 COVID-19 CT scans. An additional number of 251 CT scans were collected for training the U-Net for lobe segmentation. The evaluation of the model was done online using 60 randomly selected samples. An Offline evaluation was done using 3997 samples for five-fold cross-validation. The CT 600 samples were used (Normal, ILD, and COVID-19=200) to perform 30k iterations during the testing stage.

A weakly-supervised DL model proposed by Hu et al. [122] aimed to perform lung segmentation and COVID-19 classification. The modified VGG [123] inspires the classification and segmentation model with an increased depth, and stacked convolutional filters performed detection and classification. A ternary classification was performed between COVID-19 samples, Pneumonia, and Non-Pneumonia patients. A dataset of 450 CT samples was collected from two sources; Red Cross Society Wuhan hospital (WHRCH) [COVID-19 = 138], and Shenzhen Second Hospital (SZSH) [COVID-19 = 12, Pneumonia = 150, Non-Pneumonia = 150]. For accurate lung segmentation, a total of 60 CT were collected from three institutions with public access; Anderson Cancer Centre, Memorial Sloan-Kettering Cancer Centre, and the MAASTRO clinic.

Khan et al. [72] developed a two-stage technique: classification and segmentation of COVID-19 infected slices and regions. In this technique, COVID-19 CT images have been first classified using the new residual learning-based CoV-CTNet, and further provided to the novel CoV-RASeg models for segmentation in the second stage. These two models systematically implement region and edge-based operations to learn COVID-19 infection properties related to region homogeneity, texture variation, and boundaries. The dataset, which constituted standard CT images, was collected from SIRM [91]. A total of 370 infected samples and 459 uninfected samples were used for the classification task. In contrast, the 370 infected samples and their binary labels are also used to train the segmentation model. Moreover, the effectiveness of the developed framework has also been evaluated with MCC, Boundary F-score, pixel segmentation accuracies, and intersection over union (IOU). Wu et al. [73] presented a multi-task custom DL-based fusion model to identify COVID-19 infected patients from other Pneumonia patients. The DL-based fusion model, first extracts the lung region in each CT image via morphological optimization algorithms and threshold segmentation. The CT's Coronal, Axial, and Sagittal images are then provided to distinct blocks trained using ResNet50. The outputs of the three ResNet50 blocks are aggregated and fed to a fully-connected dense layer, which gives the risk measure of COVID-19. The dataset was gathered from three hospitals in China; Wuhan University's Renmin Hospital (COVID-19 = 265, Pneumonia = 35), Medical University's First Hospital in China (COVID-19 = 103, Pneumonia = 46), and Beijing Youan Hospital (Pneumonia = 46). A summary of the above discussion is given in Table 12.

**Table 12.** Integrated Techniques: Custom DL Models with CT

| Reference | Data Sources and Validation Scheme | Technique | Classes | Performance |
|---|---|---|---|---|
| Amyar et al.[124] | Henri Becquerel Cancer Center, Github, Public-Data (80:20%) Hold-out CV | Multi-task Learning (MTL)-based on encoder-decoder and MLP | Ternary | ACC = 94.67%, SEN = 96.0%, SPEC = 92.0%, AUC = 97.0%, DS = 88.0%, ACC=95.23%, SEN=90.2%, SPEC=99.7% |
| Wang et al.[125] | Numerous Hospitals (80:20) 5-fold CV | PARL Multi task network | Ternary | ACC = 93.3%, AUC = 97.3% F1 = 87.8% SEN=87.6%, SPEC=,95.5%, PRE=88.4% |
| Hu et al.[122] | Hospitals in China (WHRCH, SZSH) (60:20:20%) 5-fold CV | VGG inspired model | Ternary Binary | Classification: ACC = 96.2%, PRE = 97.3%, SEN = 94.5%, SPEC = 95.3%, AUC-ROCs = 97.0%. ACC=96.2%, SEN=94.5%, PRE=97.3%, SPEC=95.3%, AUC-ROCs=97.0% Segmentation: DS = 90.0% |
| Khan et al. [72] | SIRM [88] (80:20%) Hold-out CV | CoV-CTNet CoV-RASeg | Binary | Classification: ACC = 98.8%, SEN=0.99, SPEC=0.99, PRE=0.99, MCC=0.98, F1-score = 99.0%, Segmentation: DS= 95.2%, IOU = 98.7% |
| Wu et al. [73] | RHWU, FHMU, BYH (80:10:10%) Hold-out CV | ResNet50 based deep fusion model | Binary | ACC = 76%, AUC-ROC = 81.9%, SEN=81.1%,SPEC=61.5%. |

## 7  DL-based Diagnostic Applications for COVID-19 Detection

Several COVID-19 based commercial and non-commercial applications deal with diagnosis and prognosis of COVID-19 using radiological images. Tables 13 and 14 highlight these applications, while giving a brief methodology of each proposed tool. These applications aim to assist doctors and COVID-19 suspected patients by providing early interventions for disease identification. Most of these Artificial Intelligence based tools employed X-Ray imaging modality, while a few have used CT scans for COVID-19 diagnosis and prognosis.

## 8  Challenges

In recent years, many DL methods for COVID-19 diagnosis have emerged. Most of these methods used CNN for screening and analysis of COVID-19. Almost all of these methods faced some challenges which are discussed below.

## 8.1. Unavailability of Standard Radiographic Dataset

CNNs contain numerous optimization parameters; therefore, they heavily rely on large annotated datasets to produce accurate results. Since COVID-19 is a new disease therefore, a limited number of radiographic images is available in standardized format for training and testing of CNN models. The lack of data labelling facility due to shortage of radiologists in emergency is one of the main reasons of nonavailability of dataset, in addition to imaging facilities in hospitals and ethical concerns. A large amount of labeled data is required for the supervised deep CNNs. Data labelling in proper format is required to train the model, especially for segmentation tasks. Segmentation task requires large scale annotated dataset which is very difficult to acquire because labeling the 3D CT scans requires an expert radiologist that might be time consuming and costly and annotating the COVID-19 lesion. Non-availability of enough dataset hampers the training of deep learning models.

## 8.2. Variation in the Visual Quality of Images

Another important factor in CNN training is image quality. Radiographic images acquired from different sources show machine induced noise, variation in illumination, and resolution. Some of the samples show center-specific labels that induce bias during model training and affect the generalization of the model.

## 8.3. High Class Imbalance

The shortage of COVID-19 data results in another challenge which is data imbalance. The number of people affected with COVID-19 pneumonia are much lower than the number of people who suffered pneumonia caused by other kinds of viruses. Many researchers face this issue, which can be observed in [40], [43], [45], [47], [56], [70], [119]. To overcome this challenge, many techniques have been used like stratified sampling [41], data augmentation [43], [53], random under-sampling [46], binary resampling [58] and class weights [56].

## 8.4. High Computational Cost and Limited Resources

Deep CNNs are computationally expensive and require enough computational resources for training. However, lack of facilities impedes the research and development of deep learning models.

## 8.5. Optimization of DL Methods

The results of the DL models can differ based on the value of hypermeters. These hypermeters, besides others, include depth of model, choice of optimizer and loss function as well as pre-processing steps applied on images. Many researchers exploited multiple existing deep CNNs without adopting proper hyperparameter optimization strategy for COVID-19 detection thus resulting in small differences with the results of various architectures.

## 8.6. Cross-Validation Strategy for Model Evaluation

Finding a generalized of deep CNNs is a challenging task in the development of diagnostic systems. Successful clinical deployment requires rigorous validation. Most of the published papers draw training, validation and test samples from the same dataset. However, such models show poor generalization on datasets from other sources. Shortage of time is the main concern in a pandemic therefore, many models are not properly tested. Many researchers collected data from multiple online resources that might include duplicate radiographic images. Most often, the same image may occur multiple times in both training and testing, thus resulting in overfitting or misleading results. It is very crucial to detect the amount of duplication in the training and test sets to avoid data overuse. Additionally, comparison of different techniques is not possible due to differences in data distribution and division of samples.

## 8.7. Generalization of DL Models

Good generalization and optimal performance in a clinical setting normally requires the collection of datasets from multiple resources. However, collection of data from different domains is difficult due to privacy concerns, and data integration issues. Each dataset exhibits specific acquisition features depending upon the equipment and protocol used in clinical settings. Therefore, integration from multiple sources requires taking care of hardware specifications for image acquisition, visual quality and image storage format. Limited amount of data and low coverage of different COVID-19 variants make most of the models less effective on external test sets. Likewise, models trained on specific demographic regions show poor generalization towards other regions of the world.

## 8.8. Translation of Deep Learning Techniques into Clinical Setup

AI based diagnostic systems include the phases of model development, deployment and update. However, translation of diagnostic systems in clinical setup is relatively slow. Most of the efforts focused on the development of a new model with less focus on the deployment. Limited access to clinical data, interoperability with clinical settings, legal and ethical concerns are the main factors that impede the translation process.

## 8.9. Radiologists' Reservations on CAD system

Most of the published techniques lack transparency and do not define the mechanism behind decision making. The classification of radiographic images without indicating the reason for the decision may raise concerns for radiologists. Few of the studies have provided grad-cam or saliency maps that highlighted the regions considered while drawing decisions.

## 8.10. Lack of Radiological Interpretation in DL Techniques

Most of the reported works offer a limited contribution in the area of deep learning. A substantial number of papers exploited off-the-shelf transfer learning techniques for COVID-19 diagnosis, whereby the CNNs were initially designed for natural image classification. However, radiographic images are substantially different from natural images. Contrary to natural images' object, COVID-19 lesions vary in size and exhibit minor contrast variation between lesions or infected regions.

## 8.11. Lack of Interoperability Standards

DL based diagnostic systems are developed under different platforms using multiple frameworks and libraries. The deployment of these DL systems at hospitals requires standard operating procedures to ensure interoperability between hospital machinery and diagnostic systems. This requires conversion of both deep learning models and radiographic images in specific formats and hardware specific optimizations.

**Table 13.** COVID-19 Diagnosis Software Tools Using X-Rays

| Product Name | Overview | Source |
|---|---|---|
| *CAD4COVID-X-Ray* | An AI-based tool for screening of COVID-19 infection via CX-Ray. COVID-19 related deformities are detected, and a score between 0 and 100 is generated. Regions depicting any COVID-19 related abnormalities are quantified via a heat map. | https://thirona.eu/cad4covid/ |
| *qXR-Chest X-Rays* | qXR is an AI-based tool that performs screening of CX-Rays for COVID-19, Tuberculosis, and other abnormalities. It provides detection, identification, and localization of common abnormalities. | https://qure.ai/covid.html |
| *FastRAi* | This AI-based tool enhances the readability of standard X-Rays. It increases the extent of the information obtained from X-Rays by projecting them onto CT and indicates the rate of disease progression. | https://eithealth.eu/project/fastrai/ |
| *COV-Raid* | This AI-based software inputs an X-Ray image and provides COVID-19 detection and localization via a heat map. | https://covraid.com/ |
| *ATMAN* | Developed with deep convolutional neural network back-end, this AI-based tool claims to detect COVID-19 infected CX-Rays accurately despite the limited sample images' availability. | https://www.drdo.gov.in/ai-based-intelligent-covid-19-detector-technology-medical-assistance-atman |
| *X-Ray Setu* | This AI-based chatbot helps doctors in early interpretations of patient X-Rays via WhatsApp. The doctor sends a CX-Ray of the suspected patient to Setu Chatbot on WhatsApp, which responds with a report with COVID-19 infected regions marked on lungs. | https://www.xraysetu.com/ |

**Table 14.** COVID-19 Diagnosis Software Tools Using CT Scans

| Product Name | Overview | Source |
|---|---|---|
| *CAD4COVID-CT* | CAD4COVID-CT uses pulmonary lobe segmentation and generates a score between 0-5, indicating the severity of COVID-19 related abnormality for each lobe. | https://thirona.eu/cad4covid/ |
| *Covid-19 Diagnostic Assistant* | This application is available free of cost and uses an AI algorithm to analyze the CT scans and identify consistent COVID disease symptoms. The report can be accessed after scanning the generated QR code. | https://www.novarad.net/covidai |
| *BioMind Covid-19* | Using CT scans as an imaging modality, this AI-based tool differentiates between other types of Pneumonia and COVID-19 Pneumonia. It performs an assessment of lung lobes and aids in the evaluation of disease progression. | https://www.biomind.ai/product/#pneumonia |

## 9  Research Gap and Future Direction

The metrics for evaluating the efficacy of the developed models for COVID-19 diagnosis are listed in Table 1. Different assessment measures are employed across these research works, making it hard to give a comprehensive comparison when looking at the statistical analysis of COVID-19 diagnosis. While this complicates comparing the model's performance to that of other models, the use of alternative assessment criteria does not diminish the significance of these models and their outcomes. DL approaches should be investigated for additional COVID-19-related activities such as infected person tracking, COVID-19 virus mutation tracking, vaccine efficacy, and training robots to deal with the infected instances.

COVID-19's rapid appearance and contagious nature have made it difficult to obtain a large dataset for CNN training in a systematic way. In order to train the DL techniques, researchers used images taken directly from patients with severe COVID-19 or other forms of Pneumonia. However, in the actual world, a greater number of people are not afflicted by Pneumonia. The scarcity of data imposes a constraint on the ability to produce effective results. Due to this data shortage, the model can overfit and may not provide sufficiently generalized results. In this scenario, generative adversarial networks should be utilized as an effective data augmentation technique for medical images to overcome the data scarcity problem. The absence of diagnostic medical imaging equipment across all of the diagnostic facilities was one of the major limitations of the imaging modality in the diagnosis of COVID-19.

Furthermore, there is a risk of the COVID-19 virus being transferred from one patient to another due to the CT scanner tunnel contamination. As a result, the researcher should rely on X-Rays to diagnose COVID-19. The majority of COVID-19 radiography data collections are kept in a range of formats, protocols, sizes, and quality standards, providing difficulties for scientists aiming to speed up the development of COVID-19-related artificial intelligence research. Thus, for the future expansion of COVID-19 radiography data, standard operating methods should be created to facilitate academics, scientists, and anybody else with a drive and a motivation to participate and use the information without any restrictions.

Although several solutions for analyzing the COVID-19 virus utilizing DL approaches have been developed, the real-world application of these methods remains restricted and requires further refinement. With the alarming number of fatalities and affected patients being discovered every day and the virus's mutations occurring rapidly, the utilization AI on radiographic data to detect COVID-19 is still not widely implemented. In order to learn and collect information, AI techniques require a massive quantity of data and a range of computational models and CNN algorithms. The data now available to academics via open-source websites is inadequate. With the existing data, it's not ideal since the data by itself cannot describe the entire picture of the pandemic. Thus, for future research and development, specialists in AI, DL, imaging, medical, and radiology should collaborate to create reliable, accurate, global, fully accessible data and conduct research to develop different systems that can be implemented in real-time and are effective in combating this global pandemic.

**Architectural ideas for detecting Omicron**

The omicron, a new variant of COVID-19, has been discovered in 171 countries as of January 20, 2022 [126]. It is observed that the omicron has quickly surpassed the Delta variant in most nations because it has a faster growth rate. Omicron replicated 70 times faster in the human bronchi--connecting the trachea to the lungs--but 10 times slower in the lung tissues as compared to Delta and COVID-19 variants. Moreover, omicron could be ten times more infectious than the primary virus or twice as virulent as the Delta variant. Omicron showed mild symptoms against vaccinated novel SARS-CoV-2 virus patients. Unvaccinated people have a 10-fold higher risk of death than vaccinated [127]. Unvaccinated or infected patients with disabilities like chronic illnesses are more likely to develop severe symptoms and die due to an Omicron infection.

Omicron is proven to cause milder symptoms across many infected people and less severe illnesses. Omicron is likely to affect the upper respiratory system and bronchus tissue, including the nose, sinuses, throat, etc., and appear to have less severe lung infections than the Delta variant. However, COVID-19 affected both the upper and lower respiratory tracts. Upper respiratory symptoms are common in mild sickness or early SARS-CoV-2 conditions. The infection and inflammation of the lungs are common symptoms of severe disease caused by SARS-CoV-2 and prior variations [128].

Nowadays, omicron is commonly analyzed through PCR and antigen-detection rapid diagnostic tests (Ag-RDT) [129]. These screening tests may be helpful proxy markers of omicron and need to be verified for a specific setting. Moreover, using standard PCR kits, S-gene target failure (SGTF) is a proxy marker for omicron variation and can identify most Omicron strains. However, most of the omicron variant sequences have a deletion in the S gene, generating SGTF in various PCR assays [130]. Since a growing number of publicly available sequences (including all BA.2 sub-lineage sequences) miss this deletion [131]. Therefore, screening for Omicron lineages using SGTF or another manual clinical proxy marker might be missed because they lack S gene deletion. Moreover, scientists worldwide are concerned about the new variant's large number of mutations on the protein spike, making it highly contagious [132]. In this regard, the development of the automatic system is required to improve the performance of diagnostic tests for the screening of omicron and reduce the clinical burden.

This study may help in designing CNN architectures by considering infectious patterns of the new variants of COVID-19 [133]. COVID-19 infection patterns are commonly characterized by different types of opacities: ground-glass opacities (GGO), reticulation, mixed attenuation (GGO and consolidation) and pleural. These patterns include region homogeneity, structure obstruction, texture, and contrast variation. The new variants: delta and omicron, may have some common characteristics related to COVID-19. Therefore, the following deep CNN architectural ideas may be adapted to diagnose a new variant of COVID-19 infection like delta and omicron. These architectural ideas [134]–[136] can learn the infection patterns of the new variant of coronavirus that are different from other established deformed and healthy regions.

- Smoothing and edge operations and residual learning may help to capture region-homogeneity, local structural obstruction, and minor contrast variation, respectively.

- Dilated convolutional operation in multi-path transformation (split-transform-merge) extract global receptive fields and diverse feature set to learn the global structure and texture variations. Moreover, the spatial exploitation by employing small kernel sizes, symmetric and asymmetric, extracts minor contrast variation and local features.

- Implementation of channel squeezing and boosting at various levels; both abstract and target level can capture minor-contrast and texture variation of infected region.

## 10 Conclusion

COVID-19 is still an ongoing pandemic and continuing to affect the whole world. Radiological image analysis is considered a quick tool for screening. However, radiological images only demonstrate partial information about COVID-19 infected patients. Therefore, integrating the clinical examination of radiological images with DL techniques can enable us to collect knowledge from different sources and thus help in designing an accurate detection and diagnosis system. DL-based radiological image (X-Ray, CT) analysis plays a vital role in quick and accurate diagnosis. Therefore, we provide a detailed survey of DL approaches based on image and region-level analysis of COVID-19 infection. The taxonomy of the survey paper provides the effectiveness of classification, segmentation, and multi-stages approaches for detecting and diagnosing COVID-19 infected radiological images. We also give an overview of each study by describing the dataset, the number of classes, dataset distribution, partitioning, model structure, and performance evaluation criteria. Moreover, this paper highlights most of the available commercial and non- commercial diagnostic tools, dataset resources and challenges faced in the pandemic. We also discuss the possible solutions to help the AI research community for further innovations. This study may thus provide important insights for DL and medical radiological images to help researchers develop a standardized and comprehensive system for detection and analysis of COVID-19. Consequently, may help medical experts and radiologists in identifying and tackling variants of COVID-19 such as Omicron, as well as future pandemics.